\documentclass[12pt,letterpaper]{paper}
\usepackage[margin=1in]{geometry}
\usepackage[utf8]{inputenc}
\usepackage{amsmath,amssymb}
\usepackage{color,soul} 
\DeclareMathOperator{\E}{\mathbb{E}}
\DeclareMathOperator*{\Var}{\mathbb{V}ar}
\DeclareMathOperator*{\Cov}{\mathbb{C}ov}
\newcommand{\indep}{\raisebox{0.05em}{\rotatebox[origin=c]{90}{$\models$}}}
\allowdisplaybreaks
\usepackage{graphicx}
\usepackage{tikz}
\usetikzlibrary{arrows.meta}
\usepackage[font=scriptsize,labelfont=bf]{caption}
\usepackage[font=scriptsize,labelfont=bf]{subcaption} 
\usepackage{enumitem}   
\usepackage{pdflscape}
\usepackage{indentfirst} 
\usepackage{color, colortbl} 
\usepackage{array} 
\usepackage{siunitx} 
\usepackage[flushleft]{threeparttable} 
\usepackage{afterpage} 
\usepackage{setspace} 
\usepackage{ragged2e} 

\usepackage{authblk}

\begin{document}
\setlength\RaggedRightParindent{\parindent} 
\RaggedRight 
\title{\centering Sensitivity analysis for bias due to a misclassfied confounding variable in marginal structural models}
\subtitle{\centering Original Research Article}
\author[1, *]{Linda Nab}
\author[1, 2]{Rolf H.H. Groenwold}
\author[1]{Maarten van Smeden}
\author[3]{Ruth H. Keogh}
\affil[1]{Department of Clinical Epidemiology, Leiden University Medical Center, Leiden, the Netherlands}
\affil[2]{Department of Biomedical Data Sciences, Leiden University Medical Center, Leiden, the Netherlands}
\affil[3]{Department of Medical Statistics,
London School of Hygiene and Tropical Medicine, London, United Kingdom}
\affil[*]{Corresponding author: Postzone C7-P,
P.O. Box 9600, 2300 RC Leiden,
the Netherlands, l.nab@lumc.nl, tel: +31 71 526 5640 \authorcr
\vspace{2cm}
\noindent \textbf{Running head:} Sensitivity analysis for bias due to a misclassified confounding variable

\noindent \textbf{Conflicts of interest:} Non declared

\noindent \textbf{Funding:} LN, RHHG and MvS were supported by grants from the Netherlands Organization for Scientific Research (ZonMW-Vidi project 917.16.430) and Leiden University Medical Center, RHK was supported by a Medical Research Council Methodology Fellowship (MR/M014827/1) and a UK Research and Innovation Future Leaders Fellowship (MR/S017968/1). 

\noindent \textbf{Data and code availability:} The data and code used for the simulation study and sensitivity analysis have been made publicly and can be accessed via www.github.com/LindaNab/memsm.

\noindent \textbf{Acknowledgements:} We acknowledge support from Stichting Jo Kolk Studiefonds and Leids Universiteits Fonds in the form of a travel grant.}

\maketitle
\thispagestyle{empty}

\newpage
\begin{abstract} In observational research treatment effects, the average treatment effect (ATE) estimator may be biased if a confounding variable is misclassified. We discuss the impact of classification error in a dichotomous confounding variable in analyses using marginal structural models estimated using inverse probability weighting (MSMs-IPW) and compare this with its impact in conditional regression models, focusing on a point-treatment study with a continuous outcome. Expressions were derived for the bias in the ATE estimator from a MSM-IPW and conditional model by using the potential outcome framework. Based on these expressions, we propose a sensitivity analysis to investigate and quantify the bias due to classification error in a confounding variable in MSMs-IPW. Compared to bias in the ATE estimator from a conditional model, the bias in MSM-IPW can be dissimilar in magnitude but the bias will always be equal in sign. A simulation study was conducted to study the finite sample performance of MSMs-IPW and conditional models if a confounding variable is misclassified. Simulation results showed that confidence intervals of the treatment effect obtained from MSM-IPW are generally wider and coverage of the true treatment effect is higher compared to a conditional model, ranging from over coverage if there is no classification error to smaller under coverage when there is classification error. The use of the bias expressions to inform a sensitivity analysis was demonstrated in a study of blood pressure lowering therapy. It is important to consider the potential impact of classification error in a confounding variable in studies of treatment effects and a sensitivity analysis provides an opportunity to quantify the impact of such errors on causal conclusions. An online tool for sensitivity analyses was developed: https://lindanab.shinyapps.io/SensitivityAnalysis. 
\end{abstract}
(Keywords: marginal structural models, inverse probability weighting, misclassification, point-treatment study, sensitivity analysis)

\section{Introduction}
The aim of many observational epidemiologic studies is to estimate a causal relation between an exposure and an outcome. One of the fundamental challenges in making inference about causal effects from observational data is adequately dealing with confounding. In the case of a point-treatment, that is estimating the effect of a treatment at a single time point on a subsequent outcome, many methods exist that aim to estimate average treatment effects (ATEs). These include traditional conditional regression analysis as well as marginal structural models estimated using inverse probability weighting (MSMs-IPW) \cite{Hernan2002EstimatingMeasures,Robins2000MarginalEpidemiology}. Unlike conditional regression, MSMs extend more easily to longitudinal settings with time-dependent confounding.

To obtain valid inference, MSMs-IPW, like other methods to correct for confounding, assume that confounding variables are measured without error, an assumption that is hardly ever warranted in observational epidemiologic research \cite{Rubin2008ForAnalysis,Steiner2011OnScores,Michels2001AError,vanSmeden2019ReflectionResearch}. For instance, CD4 count is known to be error-prone \cite{Raboud1995QuantificationIndividuals}, but used as a time-dependent confounder in one of the introductory papers of MSMs, studying the effect of zidovudine therapy on mean CD4 count in HIV-infected men \cite{Hernan2002EstimatingMeasures}. Another example of the use of an error-prone confounding variable is $\gamma$-glutamyltransferase in investigating the relationship between Hepatitus C virus treatment and progression of liver disease \cite{Kyle2016CorrectingModels}. In both examples, the assumption of measurement error-free confounding variables is possibly violated and may lead to bias in the treatment effect estimator.

There is a substantial literature on bias due to measurement error in confounding variables in conditional regression analyses \cite{Buonaccorsi2010MeasurementApplications,Carroll2006,Gustafson2004MeasurementAdjustments.,Fuller1987MeasurementModels}, but the impact of measurement error in confounding variables in causal inference methods, such as MSMs-IPW, has not received much attention. One exception is a study by Regier et al. that showed by means of a simulation study that measurement error in continuous confounding variables can introduce bias in the ATE in a point-treatment study \cite{Regier2014TheStudy}. McCaffrey et al. proposed a weighting method to restore the treatment effect estimator when covariates are measured with error \cite{McCaffrey2013InverseCovariates}. 

We provide a discussion of when measurement error in a confounding variable is important for estimating treatment effects and when it is not. In addition, we derive expressions that quantify the bias in the ATE if a dichotomous confounding variable is misclassified, focusing on a point-treatment study with a continuous outcome. These expressions allow us 1) to quantify the bias due to classification error in a confounding variable in MSMs-IPW, and to compare this with the bias from a conditional regression analysis and 2) to inform sensitivity analyses to assess the uncertainty of study results if the assumption of no classification error in a confounding variable is violated \cite{Greenland1996BasicBiases, Lash2009ApplyingData, Lash2014GoodAnalysis}. Simulation results are used to study the finite sample performance of a MSM-IPW and compared to that of conditional regression models if classification error in a confounding variable is present. Finally, we illustrate our sensitivity analysis in a study of the effect of blood pressure lowering drugs on blood pressure.

\begin{singlespacing}
\section{Settings and impact of measurement error, notation and assumptions}\label{sec:covme}
\end{singlespacing}

Let $A$ denote the treatment indicator and $Y$ the outcome. Let there be a variable $L$ that confounds the association between treatment and outcome and suppose that, instead of confounding variable $L$, the error-prone confounding variable $L^*$ is observed. We consider two settings in which measurement error in confounding variables may occur and discuss the impact of measurement error in both settings.

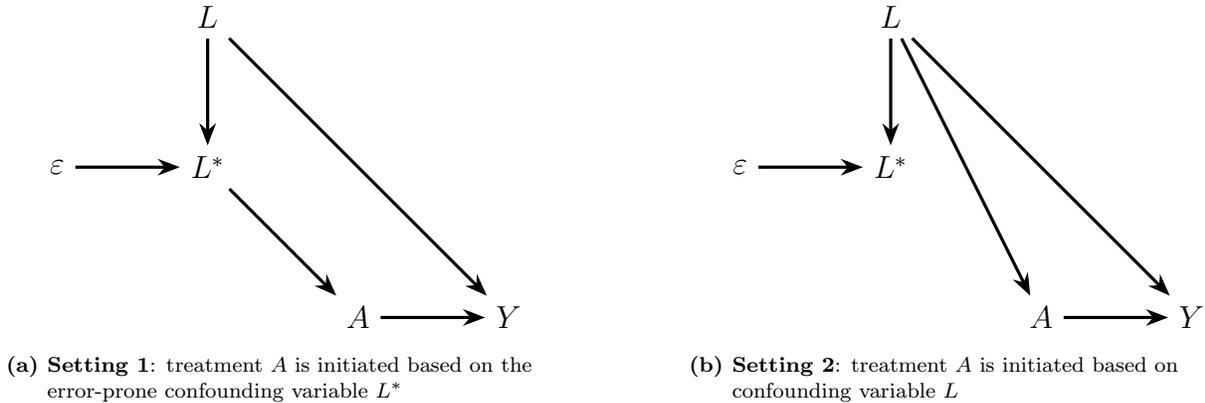
\begin{figure}[ht]
\begin{center}
\begin{subfigure}{.45\textwidth}
\centering
    \begin{tikzpicture}
    \begin{scope}
        \node (L) at (-1, 4) {$L$};
        \node (Ls) at (-1,2) {$L^*$};
        \node (eps) at (-3,2) {$\varepsilon$};
        \node (A) at (1,0) {$A$};
        \node (Y) at (3,0) {$Y$};
    \end{scope}
    \begin{scope}[>={Stealth[black]}, every edge/.style={draw=black, very thick}]
        \path [->] (L) edge node {} (Ls);
        \path [->] (A) edge node {} (Y);
        \path [->] (Ls) edge node {} (A);
        \path [->] (L) edge node {} (Y);
        \path [->] (eps) edge node {} (Ls);
    \end{scope}
    \end{tikzpicture}
    \caption{\textbf{Setting 1}: treatment $A$ is initiated based on the error-prone confounding variable $L^*$}\label{fig:dags:a}
\end{subfigure}\hfill
\begin{subfigure}{.45\textwidth}
\centering
    \begin{tikzpicture}
    \begin{scope}
        \node (L) at (-1, 4) {$L$};
        \node (Ls) at (-1,2) {$L^*$};
        \node (eps) at (-3,2) {$\varepsilon$};
        \node (A) at (1,0) {$A$};
        \node (Y) at (3,0) {$Y$};
    \end{scope}
    \begin{scope}[>={Stealth[black]}, every edge/.style={draw=black, very thick}]
        \path [->] (L) edge node {} (Ls);
        \path [->] (A) edge node {} (Y);
        \path [->] (L) edge node {} (A);
        \path [->] (L) edge node {} (Y);
        \path [->] (eps) edge node {} (Ls);
    \end{scope}
    \end{tikzpicture}
    \caption{\textbf{Setting 2}: treatment $A$ is initiated based on confounding variable $L$}\label{fig:dags:b}
\end{subfigure}
\caption{Measurement error $\varepsilon$ in variable $L$ that confounds the association between treatment $A$ and outcome $Y$ in two settings illustrated in Directed Acyclic Graphs}\label{fig:dags}
\end{center}
\end{figure}

\textbf{Settings and impact of measurement error.} The Directed Acyclic Graph (DAG) in Figure \ref{fig:dags:a} illustrates \textbf{setting 1}. In this setting, treatment initiation is based on the error-prone confounding variable rather than the underlying truth. Consider for example a study investigating the effect of zidovudine therapy ($A$) on mean CD4 count ($Y$) in HIV-infected patients. Past CD4 count ($L$) confounds the relation between zidovudine therapy and future CD4 count \cite{Hernan2002EstimatingMeasures},  but the observed past CD4 count is prone to measurement error ($L^*$) \cite{Raboud1995QuantificationIndividuals}. Yet, the actual information that the clinician uses to initiate treatment is the observed error-prone CD4 count ($L^*$) instead of true CD4 count ($L$), as depicted in Figure \ref{fig:dags:a} (measurement error in the outcome, in our example also CD4 count, is not depicted here). In setting 1, conditioning on error-prone confounding variable $L^*$ will block the backdoor path from treatment $A$ to outcome $Y$. Thus, if there is confounding by indication but the error-prone variable is used for treatment decisions, it is sufficient to control for the error-prone confounding variable to estimate the causal effect of treatment on outcome. We recognize this as an important exception since this means that measurement error in a confounding variable will not always lead to bias.

The DAG in Figure \ref{fig:dags:b} illustrates \textbf{setting 2}, in which treatment initiation is based on confounding variable $L$, but only a proxy of $L$ is observed ($L^*$). An example here might be a study investigating the effect of influenza vaccination ($A$) on mortality ($Y$) in the elderly population \cite{Jackson2006EvidenceSeniors}. Frailty ($L$) possibly confounds the association between influenza vaccination and mortality. Frailty is observed by a clinician and may influence vaccination probability, but only a proxy of frailty ($L^*$) may be available in patient record data, as depicted in Figure \ref{fig:dags:b}. In this setting, conditioning on $L^*$ will not fully adjust for confounding by $L$, because conditioning on $L^*$ does not block the backdoor path from $A$ to $Y$ via $L$.

\textbf{Notation and assumptions.} The above is a graphical interpretation and we will now continue investigating the impact of classification error in setting 2, by focusing on the setting where that $L$ is a dichotomous confounding variable and $Y$ a continuous outcome. We use the potential outcomes framework \cite{Rubin1974EstimatingStudies, Rubin1990FormalEffects}. Let $Y^{a=0}$ denote the outcome that a individual would have had if treatment $A$ was set to $a=0$, and let $Y^{a=1}$ denote the outcome if treatment $A$ was set to $a=1$. We assume that $L^*$ is non-differentially misclassified with respect to the outcome ($L^* \indep Y|L$) and to the treatment ($L^* \indep A|L$) (the notation $Q \indep R|S$ denotes that a variable $Q$ is independent of another variable $R$ given a third variable $S$). Let $p_1$ denote the sensitivity of $L^*$ and $1-p_0$ the specificity of $L^*$ (i.e., $P(L^*|L=l)=p_l$). We also denote the probability of treatment given the level of $L$ by $P(A=1|L=l) = \pi_l$ and the prevalence of $L$ by $P(L=1) = \lambda$.

We also assume that the following causal assumptions are satisfied to recover the causal effect of treatment on the outcome. Under the \textit{consistency} assumption, we require that we observe $Y=Y^{a=0}$ if the individual is not exposed, or $Y=Y^{a=1}$ if the individual is exposed \cite{Hernan2019CausalInference}. Further, we assume that the potential outcome $Y^a$ for an individual does not depend on treatments received by other individuals, also referred to as \textit{Stable-Unit-Treatment-Value-Assumption} \cite{Rubin1980RandomizationComment}. Additionally, we assume \textit{conditional exchangeability}, i.e., given the level of the confounding variable $L$, the outcome in the untreated would have been the same as the outcome in the treated had individuals in the untreated group received treatment \cite{Hernan2019CausalInference}. In notation, $A\indep Y^{a}|L$, for $a = 0,1$. Finally, we assume $\pi_L > 0$ for $L = 0, 1$ (\textit{positivity}) \cite{Cole2008ConstructingModels}.

For causal contrasts, we compare expected potential outcomes (i.e., counterfactual outcomes) under the two different treatments. The average causal effect of the treatment on the outcome is $\beta=\E[Y^{a=1}]-\E[Y^{a=0}]$. Under the above defined assumptions, the conditional effect of treatment $A$ on outcome $Y$ can be defined through the following linear model:
\begin{equation}\label{eq:lm}
    \E[Y^a|L]=\E[Y|A=a,L]=\alpha+\beta a + \gamma L.
\end{equation}
Estimates for $\beta$ in the above model can be obtained by fitting a conditional regression model. Alternatively, the effect of treatment $A$ on outcome $Y$ may be estimated by fitting a MSM:
\begin{equation}\label{eq:msm}
    \E[Y^a]=\alpha_{\mathrm{msm}} + \beta a, \qquad \mathrm{where } \quad \alpha_{\mathrm{msm}} = \alpha + \gamma \E[L].
\end{equation}
Estimates for $\beta$ in the above model can be obtained by IPW estimation: by fitting a linear regression model for $Y$ on $A$ where the contribution of each individual is weighted by 1 over the probability of that individual's observed treatment given the confounding variable $L$ \cite{Robins2000MarginalEpidemiology}, estimating the marginal treatment effect. Since our focus is on linear models, the conditional and marginal treatment effect, respectively denoted by $\beta$ in model (\ref{eq:lm}) and (\ref{eq:msm}), are equal (not generally true for non-linear models \cite{Robins2000MarginalEpidemiology}).

\begin{singlespacing}
\section{Quantification of bias due to classification error in a confounding variable}\label{sec:covme:impact}
\end{singlespacing}
Our aim is to study the effect of using the misclassified confounding variable $L^*$ in place of the confounding variable $L$ in the conditional regression model or in the model for the weights used to fit the MSM on the ATE estimator in the setting where $L$, not $L^*$, influences treatment initiation (setting 2 above).

\textbf{Conditional model.} By the law of total expectation, the expected value of the outcome $Y$ given treatment $A$ and misclassified confounding variable $L^*$ is (see appendix section \ref{ap:sec:cm} for further detail),
\begin{eqnarray*}
    \E[Y|A=a, L^*] = \E_{L|A=a, L^*}\big[\E[Y|A=a, L^*, L]\big] &=& \{\alpha + \gamma \phi_{00} + \delta u_0\}\\
    &+& \{\beta + \gamma (\phi_{10} - \phi_{00}) + \delta u_A\}a \\
    &+& \{\gamma(\phi_{01}-\phi_{00}) + \delta u_{L^*}\}L^*,
\end{eqnarray*}
where $\phi_{al^*}=P(L=1|A=a, L^*=l^*)$, $\delta = \E[Y|A=1, L^*=1]=\gamma(\phi_{11}-\phi_{10}-\phi_{01}+\phi_{00})$ and $u_0, u_A, u_{L^*}$ represent the coefficients of the linear model $\E[AL^*|A,L^*]=u_0 + u_AA + u_{L^*}L^*$, modeling the mean of treatment $A$ times the misclassified confouding variable $L^*$ (i.e., $AL^*$) given $A$ and $L^*$ (see next paragraph for an explanation of why these appear). The coefficient for treatment $A$ in the above model is $\beta + \gamma (\phi_{10} - \phi_{00}) + \delta u_A$, and is therefore biased for the parameter of interest ($\beta$). By rewriting $u_A$ in terms of $\lambda$, $\pi_0$, $\pi_1$, $p_0$ and $p_1$ (see appendix section \ref{ap:sec:cm}), we find that the bias due to classification error in $L^*$ in the ATE in a conditional regression model is, 
\begin{eqnarray}\label{eq:biascm}
    \mathrm{Bias}_{\mathrm{cm}}(\beta) = \gamma(\phi_{10}-\phi_{00})\left(1-\ell\times\Big\{\frac{\pi^*_1(1-\omega)-\pi^*_1(\pi^*_1-\pi^*_0)(1-\ell)}{\omega(1-\omega) - (\pi^*_1-\pi^*_0)^2\ell(1-\ell) }\Big\}\right)\nonumber\\
+ \gamma(\phi_{11}-\phi_{01})\left(\ell\times\Big\{\frac{\pi^*_1(1-\omega)-\pi^*_1(\pi^*_1-\pi^*_0)(1-\ell)}{\omega(1-\omega) - (\pi^*_1-\pi^*_0)^2\ell(1-\ell) }\Big\}\right),
\end{eqnarray}
where $\pi^*_{l^*}=P(A=1|L^*=l^*)$, $\ell = P(L^*=1)$ and $\omega = P(A=1)$ (see the appendix section \ref{ap:sec:cm} for a derivation).

We focused above on a model for $Y$ conditional on $A$ and $L^*$ which includes only main effects of $A$ and $L^*$, as this is typically what would be done in practice when replacing $L$ with $L^*$. In fact, it can be shown that when the model for the outcome $Y$ given $A$ and $L$ includes only main effects of $A$ and $L$, the implied correctly specified model for $Y$ given $A$ and $L^*$ also includes an interaction between $A$ and $L^*$, explaining the appearance of $u_0, u_A$ and $u_L$ in the above since the interaction is not modelled. We refer to the appendix section \ref{ap:sec:cm} for the bias in the ATE in a model including an interaction between the treatment and the misclassified confouding variable.

\textbf{MSM estimated using IPW.} A MSM-IPW proceeds by fitting a linear regression for outcome $Y$ on treatment $A$ where the contribution of each individual is weighted by 1 over the probability of that individual's observed treatment given misclassified confounding variable $L^*$ \cite{Robins2000MarginalEpidemiology}. An estimator for the ATE $\beta$ is,
\begin{eqnarray*}
    \hat{\beta}=\frac{\sum_{i=1}^{n}\frac{1}{P(A_i|L_i^*)}(Y_i-\overline{Y}_{w})(A_i-\overline{A}_{w})}{\sum_{i=1}^{n}\frac{1}{P(A_i|L_i^*)}(A_i-\overline{A}_{w})^2} \quad
\text{where,} \quad
    \overline{Y}_{w} &=& \frac{\sum_{i=1}^{n}Y_i/P(A_i|L_i^*)}{\sum_{i=1}^{n}1/P(A_i|L_i^*)}\\
    \text{and,} \quad \overline{A}_{w} &=& \frac{\sum_{i=1}^{n}A_i/P(A_i|L_i^*)}{\sum_{i=1}^{n}1/P(A_i|L_i^*)}.
\end{eqnarray*}
It can be shown that $\E[\hat{\beta}] = \beta + \gamma(1-\ell)(\phi_{10} - \phi_{00}) + \gamma\ell(\phi_{11} - \phi_{01})$. Consequently, the bias in the ATE $\beta$ in a MSM estimated using IPW is,
\begin{equation}\label{eq:biasmsm}
    \mathrm{Bias}_{\mathrm{msm}}(\beta) = \gamma(\phi_{10} - \phi_{00})(1-\ell) + \gamma(\phi_{11} - \phi_{01})\ell.
\end{equation}
We refer to the appendix section \ref{ap:sec:msm} for a derivation of the above formula. 

\subsection{Exploration of bias} 
To study the bias due to misclassification from the conditional model and MSM-IPW, we analytically and graphically explore bias expressions (\ref{eq:biascm}) and (\ref{eq:biasmsm}).

\textbf{Null-bias.} From the bias expressions, it can be seen that there are four trivial conditions in which the bias in the ATE is null in both the conditional model and MSM-IPW, in line with general understanding of causal inference \cite{Steiner2016TheBiases}. 1) If there is \textit{no classification error in the observed confounding variable $L^*$}, i.e., specificity is 1 ($p_0 = 0$) and sensitivity is 1 ($p_1 = 1$), it follows that the confounding variable $L$ corresponds to the observed variable $L^*$, irrespective of treatment level (i.e., $\phi_{10}=0$, $\phi_{00}=0$, $\phi_{11}=1$ and $\phi_{01}=1$). 2) If the \textit{true relation between variable $L$ and outcome $Y$ is null} (i.e., $\gamma$ is zero, thus there is no arrow from $L$ to $Y$ in Figure (\ref{fig:dags:b})). 3) If the \textit{variable $L$ does not affect the probability of receiving treatment} (i.e., $\pi_0=\pi_1$, thus there is no arrow from $L$ to $A$ in Figure (\ref{fig:dags:b})), the probability that the variable $L$ is 1 depends on the value of the misclassified variable $L^*$ but no longer on $A$ (i.e., $\phi_{00} = \phi_{10}$ and $\phi_{01} = \phi_{11}$). 4) If the \textit{prevalence of the confounding variable $L$ is null or one} (i.e., $\lambda = 0$ or $\lambda = 1$) bias is null since if $\lambda = 0$ then the probability that $L$ is one given $A=a$ and $L^*=l^*$ is null (i.e., $\phi_{00}=\phi_{01}=\phi_{10}=\phi_{11}=0$) and if $\lambda = 1$, then $\phi_{00}=\phi_{01}=\phi_{10}=\phi_{11}=1$.

\textbf{Equal biases.} In some cases, bias in the ATE from the conditional regression analysis equal to that from the MSM-IPW. Evidently, biases are identical if bias in both models is null (see first paragraph of this section). Additionally, bias formula \ref{eq:biascm} and \ref{eq:biasmsm} show that bias from the two methods is equal if the term between curly brackets in equation (\ref{eq:biascm}) is equal to 1.

\textbf{Sign and magnitude of bias.} Figure \ref{fig:biasplots} illustrates bias in the ATE in a conditional model and a MSM-IPW, obtained by using the bias expressions and by varying either the probability of receiving treatment if the confounding variable is one (i.e., $0 < \pi_1 < 1$) or the effect of the confounding variable on the outcome (i.e., $-5 \leq \gamma \leq 5$) or the prevalence of the confounding variable (i.e., $0 \leq \lambda \leq 1$) or 1 minus the specificity or sensitivity of the error-prone confounding variable (i.e., $0 \leq p_0 \leq 1$ and $0 \leq p_1 \leq 1$, respectively) and keeping the other ones fixed.

Figures \ref{fig:biasplots:a}-\ref{fig:biasplots:d} show that the direction of the bias depends on whether the effect of the confounding variable on the outcome is positive or negative (i.e., $\gamma>0$ or $\gamma<0$, respectively) and whether the probability of receiving treatment given that confounding variable $L$ is one, is greater, equal or smaller than the probability of receiving treatment given that $L$ is null ($\pi_1:\pi_0>1$, $\pi_1 = \pi_0$ or $\pi_1:\pi_0<1$, respectively). The lines showing the bias under the two methods cross at point (1,0) in Figure \ref{fig:biasplots:a} and \ref{fig:biasplots:b} and at point (0,0) in Figure \ref{fig:biasplots:c} and \ref{fig:biasplots:d}. This, together with the fact that the functions are strictly monotonic, show that the bias in a MSM-IPW cannot be negative if the bias in the conditional model is positive and vice versa (i.e., the bias will be in the same direction for both models). Figures \ref{fig:biasplots:a} and \ref{fig:biasplots:b} show that in absolute values the bias is greatest when $\pi_0=$ and $\pi_1=1$. 

In Figure \ref{fig:biasplots:e} and \ref{fig:biasplots:f}, the prevalence of the confounding variable $L$ was varied ($\lambda$) while $\pi_1$, $\pi_0$, $\gamma$, $p_1$ and $p_0$ were kept constant. Given these variables, there is a $\lambda$ for which the bias curve has a maximum (if bias is greater than zero) or minimum (if bias is smaller than zero). The specificity and sensitivity of $L^*$ were kept constant across Figure \ref{fig:biasplots:a}-\ref{fig:biasplots:f}, i.e., the specificity was $0.95$ ($p_0 = 0.05$) and the sensitivity was $0.90$ ($p_1 = 0.90$). In Figure \ref{fig:biasplots:g}, 1 minus specificity was varied (i.e., $0 \leq p_0 \leq 1$). Bias is smallest if the specificity is one (i.e., $p_0 = 0$) and is maximal if the 1 minus specificity equals sensitivity (i.e., $p_0 = p_1$). In Figure \ref{fig:biasplots:h}, sensitivity was varied (i.e., $0 \leq p_1 \leq 1$). Bias is smallest if the sensitivity is one (i.e., $p_1 = 1$) and is maximal if the 1 minus specificity equals sensitivity (i.e., $p_0 = p_1$). An online application has been developed which can be used to obtain bias plots for other combinations of the parameters available at: https://lindanab.shinyapps.io/SensitivityAnalysis.

\afterpage{%
\begin{landscape}
\centering
\begin{figure}[hp]
\begin{subfigure}{.25\textwidth}
\centering
\includegraphics[scale = 0.4]{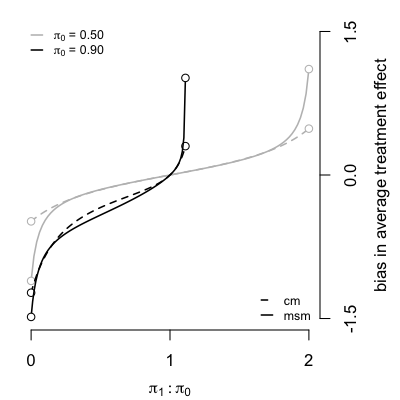}
\caption{}\label{fig:biasplots:a}
\end{subfigure}\hfill
\begin{subfigure}{.25\textwidth}
\centering
\includegraphics[scale = 0.4]{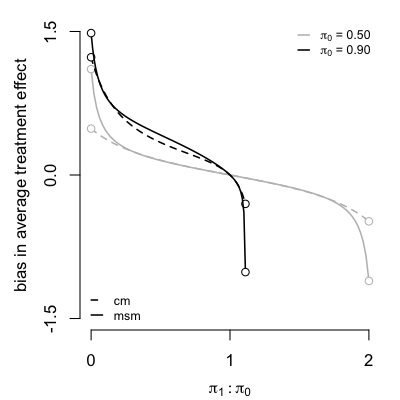}
\caption{}\label{fig:biasplots:b}
\end{subfigure}\hfill
\begin{subfigure}{.25\textwidth}
\centering
\includegraphics[scale = 0.4]{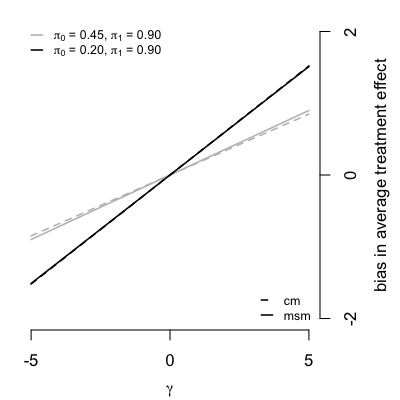}
\caption{}\label{fig:biasplots:c}
\end{subfigure}\hfill
\begin{subfigure}{.25\textwidth}
\centering
\includegraphics[scale = 0.4]{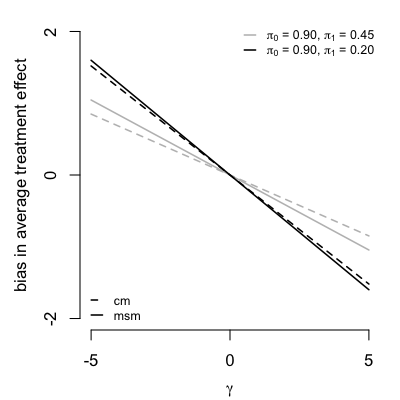}
\caption{}\label{fig:biasplots:d}
\end{subfigure}

\begin{subfigure}{.25\textwidth}
\centering
\includegraphics[scale = 0.4]{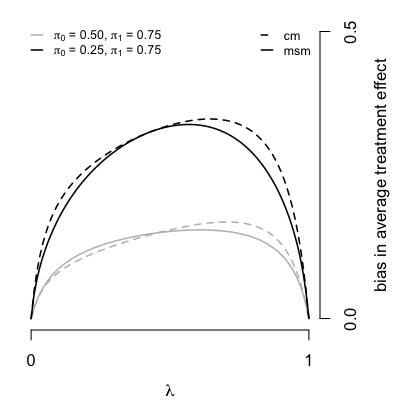}
\caption{}\label{fig:biasplots:e}
\end{subfigure}\hfill
\begin{subfigure}{.25\textwidth}
\centering
\includegraphics[scale = 0.4]{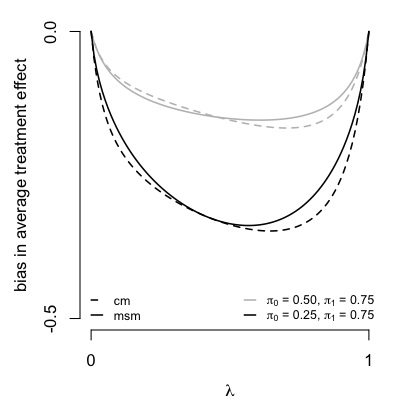}
\caption{}\label{fig:biasplots:f}
\end{subfigure}\hfill
\begin{subfigure}{.25\textwidth}
\centering
\includegraphics[scale = 0.4]{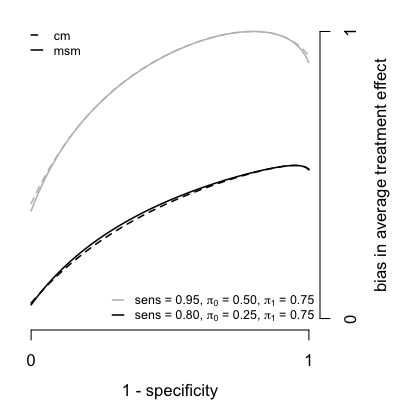}
\caption{}\label{fig:biasplots:g}
\end{subfigure}\hfill
\begin{subfigure}{.25\textwidth}
\centering
\includegraphics[scale = 0.4]{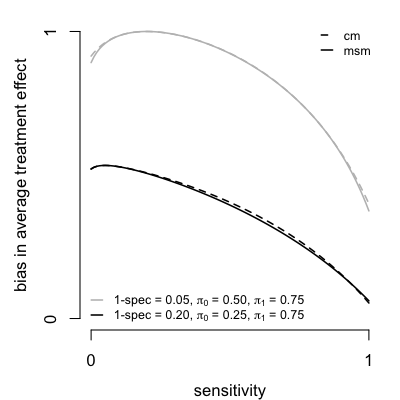}
\caption{}\label{fig:biasplots:h}
\end{subfigure}
\caption{Bias in the average treatment effect estimator in a marginal structural model estimated using inverse probability weighting (msm, solid lines) and conditional model (cm, dashed lines) under classification error in the confounding variable. In graphs a-f, the specificity of the misclassified confounding variable is 0.95 ($p_0 = 0.05$) and the sensitivity of the misclassified confounding variable is 0.9 ($p_1=0.9$). In all graphs except (e) and (f), the prevalence of the confounding variable 0.5 (i.e., $\lambda = 0.5$). Graph (a): the probability of receiving treatment if the confounding variable is 1 varies from 0 to 1 (i.e., $0 \leq \pi_1 \leq 1$) and the effect of the confounding variable on the outcome is positive ($\gamma = 2$). Graph (b): same as (a) apart from that $\gamma = -2$. The open points in graphs (a) and (b) depict that bias for these parameters is not defined as the ATE is not defined (non-positivity). Graph (c): the effect of the confounding variable on outcome is varied (i.e., $-5 \leq \gamma \leq 5$) and the probability of receiving treatment is greater if the confounding variable is 1 than if 0 (i.e., $\pi_1:\pi_0 > 1$). Graph (d): same as (c) apart from that $\pi_1:\pi_0 < 1$. Graph (e): the prevalence of the confounding variable is varied (i.e., $0 \leq \lambda \leq 1$), $\pi_1:\pi_0 > 1$ and $\gamma = 2$. Graph (f): same as (e) apart from that $\gamma = -2$. Graph (g): the specificity is varied (i.e., $0 \leq p_0 \leq 1$), $\pi_1:\pi_0 > 1$, $\gamma = 2$ and the sensitivity is 0.95 or 0.80 ($p_1 = 0.95$ or $p_1 = 0.80$). Graph (h): same as (f) apart from that the sensitivity is varied (i.e., $0 \leq p_1 \leq 1$) and the specificity is 0.95 or 0.80 ($p_0 = 0.05$ or $p_1 = 0.20$).}\label{fig:biasplots}
\end{figure}
\end{landscape}}

\subsection{Simulation study}\label{sec:simstudy}
A simulation study was conducted to study the finite sample properties of MSMs estimated using IPW and conditional models if there is classification error in the confounding variable. Five-thousand data sets were generated with sample sizes of 1,000 and 100, using the following data generating mechanisms:
\begin{eqnarray*}
L &\sim& \operatorname{Bern} \left({\lambda}\right), \quad A|L \sim \operatorname{Bern} \left({\pi_0^{(1-L)}\pi_1^L}\right),\\ L^*|L &\sim& \operatorname{Bern} \left({p_0^{(1-L)}p_1^L}\right) \quad \text{and} \quad
Y|A,L \sim  \mathcal{N}(1 + \beta A + \gamma L, 1).
\end{eqnarray*}
Five different scenarios were studied, of which the parameters values can be found in Table \ref{tab:simscen}. In all scenarios, the ATE $\beta$ (estimand) is 1 and the association between the confounding variable $L$ and outcome $Y$ is positive (i.e., $\gamma = 2$). In scenario 0, we assume no classification error. In scenarios 1-4, we assume that the error-prone variable $L^*$ is a misclassified representation of $L$ with a sensitivity of 0.95 (i.e., $p_0 = 0.05$) and a specificity of 0.90 (i.e., $p_1 = 0.9$). In scenario 1, bias in the ATE $\beta$ is expected to be negative since the probability of receiving treatment given that $L$ is not present is greater than receiving treatment given that $L$ is present, and the association between $L$ and $Y$ is positive (i.e., $\pi_0 > \pi_1$ and $\gamma = 2$). In contrast, in scenario 2 and 3, bias in the ATE is expected greater than null, since $\pi_0 < \pi_1$ and $\gamma = 2$. Further, after investigation of Figure \ref{fig:biasplots}, we expect that bias in the ATE estimated in the conditional model is greater than bias in the ATE in a MSM estimated using IPW in scenario 2 (and vice versa in scenario 3 and 4). Finally, in scenario 4, we expect that bias in the ATE from the conditional model is equal to that in a MSM estimated using IPW.

\textbf{Model estimation and performance measures.} The ATE $\beta$ (estimand) was obtained by fitting a conditional model using conditional regression and by fitting a MSM estimated using IPW, both using the misclassified confounding variable $L^*$ instead of the confounding variable $L$ from the data generating mechanism. For the MSM analysis we used the R package \texttt{ipw} \cite{RCoreTeam2018R:Computing} \cite{VanDerWal2011Ipw:Weighting}. Performance of both models was evaluated in terms of the bias, mean model-based standard error standard error, and mean square error of the estimated treatment effect, the percentages of 95\% confidence intervals that contain the true value of the estimand (coverage), the empirical standard deviation of the estimated treatment effects (empSE) and mean model based standard error of the estimated treatment effect. Robust model based standard errors of the ATE in a MSM estimated using IPW were estimated using the R package \texttt{survey} \cite{Lumley2004AnalysisSamples}. Monte Carlo standard errors were calculated for all performance measures \cite{Morris2019UsingMethods}, using the R package \texttt{rsimsum} \cite{Gasparini2018Rsimsum:Studies}. Additionally, theoretical bias of the ATE in both methods was calculated based on the bias expressions defined by model (\ref{eq:biascm}) and model (\ref{eq:biasmsm}).

\textbf{Results.} Table \ref{tab:ressim} shows the results of the simulation study. Bias found in the simulation study corresponds to the theoretical bias derived from the bias expressions. The standard deviation of the ATE estimates (empSE) from the MSM estimated using IPW is equal to or greater than that from the conditional model. Yet, in the scenarios where bias in the ATE in the MSM estimated using IPW was smaller than bias in the conditional model (scenarios 2 and 3), empSE of both methods was equal, and hence, MSE is smaller for one method if also bias is smaller. Furthermore, the (robust) model based standard errors of the ATE in a MSM estimated using IPW are conservative and greater than the empirical standard errors, since the uncertainty in estimating the treatment weights is not taken into account. Allowing for the estimation of the weights will shrink the standard errors \cite{Robins2000MarginalEpidemiology, Robins2000MarginalInference}. Consequently, confidence intervals of the treatment effect obtained in a MSM estimated using IPW are generally wider and coverage of the true treatment effect is higher compared to a conditional model, ranging from over coverage if there is no classification error to smaller under coverage when there is classification error.

\begin{singlespacing}
\section{Illustration: sensitivity analysis of classification error in a confounding variable}
\end{singlespacing}
Sensitivity analyses provide a tool to incorporate uncertainty in study results due to systematic errors \cite{Greenland1996BasicBiases,Lash2014GoodAnalysis}. Using an example study of blood pressure lowering therapy, we will illustrate how the bias expressions in section \ref{sec:covme:impact} can be used to perform a sensitivity analysis for misclassification in a confounder.

\textbf{Application.}
To illustrate how the bias expressions can be used in a sensitivity analysis for the ATE, we use data of the National Health And Nutritional Examination Survey (NHANES) \cite{Centershttps://wwwn.cdc.gov/nchs/nhanes/continuousnhanes/default.aspxBeginYear=2011., Centershttps://wwwn.cdc.gov/nchs/nhanes/continuousnhanes/default.aspxBeginYear=2013.}. Specifically, we study the effect of diuretic use ($A=1$) in comparison to beta blocker use $(A=0$) on systolic blood pressure ($Y$), adjusted for self-reported categorical body mass index (BMI) ($L^*$). For this illustration, we categorise self-reported BMI into two distinct categories: underweight/normal weight (BMI $<25$ ($L^*=0$)) and overweight/obese (BMI $\geq 25$ ($L^*=1$)). However, we stress that one should preferably not categorise BMI in most practical applications \cite{Altman2006TheVariables}. 

We assume that blood pressure lowering therapy is initiated based on the true BMI ($L$) instead of the observed self-reported BMI available in our data (setting 2, depicted in Figure \ref{fig:dags:b}). Further, we consider BMI the only confounding variable, which is a simplification of reality. Our earlier results indicate that if we use self-reported measures to adjust for BMI instead of the true level of BMI, the ATE will be biased. To quantify how large the bias in the ATE is expected to be due to classification error in self-reported BMI category, we perform a sensitivity analysis using the bias expressions presented in section \ref{sec:covme:impact}.

The NHANES survey consists of questionnaires, followed by a standardized health examination in specially equipped mobile examination centers. In the 2011-2014 sample 19,151 participants were physically examined. Of the 19,151 physically examined people, 12,185 participants aged over 16 were asked to fill out a questionnaire, including questions on self-reported weight and height, used to calculate self-reported BMI. For this illustration, we used complete data on 585 users of diuretics and 824 users of beta blockers (excluding non-users and people using both).

\textbf{Average treatment effect.} Table \ref{tab:resultsnhanes} shows the ATE of diuretics use in comparison to beta blocker use on mean systolic blood pressure, unadjusted and adjusted for self-reported BMI category. In a MSM estimated using IPW, an ATE (95 \% CI) of $-3.52$ $(-1.21; -5.74)$ was found. In a conditional regression model, the ATE (95 \% CI) was found to be $-3.48$ $(-1.27; -5.76)$.

\textbf{Sensitivity analysis} To inform the bias expressions, we need to make assumptions on the sensitivity and specificity of the self-reported BMI as well as that classification errors are non-differential with respect to blood pressure and treatment. We speculate reasonable ranges for the sensitivity and specificity of self-reported BMI category of $0.90$ to $0.98$. Reports in the literature and/or a researcher's own experience should inform these parameters. Further, in the NHANES data, it was found that the prevalence of self-reported overweight/obese was 0.77 ($\ell$), the probability of receiving treatment given that one self-reports to be underweight/normal weight is 0.32 ($\pi^*_0$), the probability of receiving treatment given that one self-reports to be overweight/obese is 0.44 ($\pi^*_1$). The prevalence of diuretic treatment use was 0.42 ($\omega$).

\begin{figure}
    \centering
    \includegraphics[scale = 0.75]{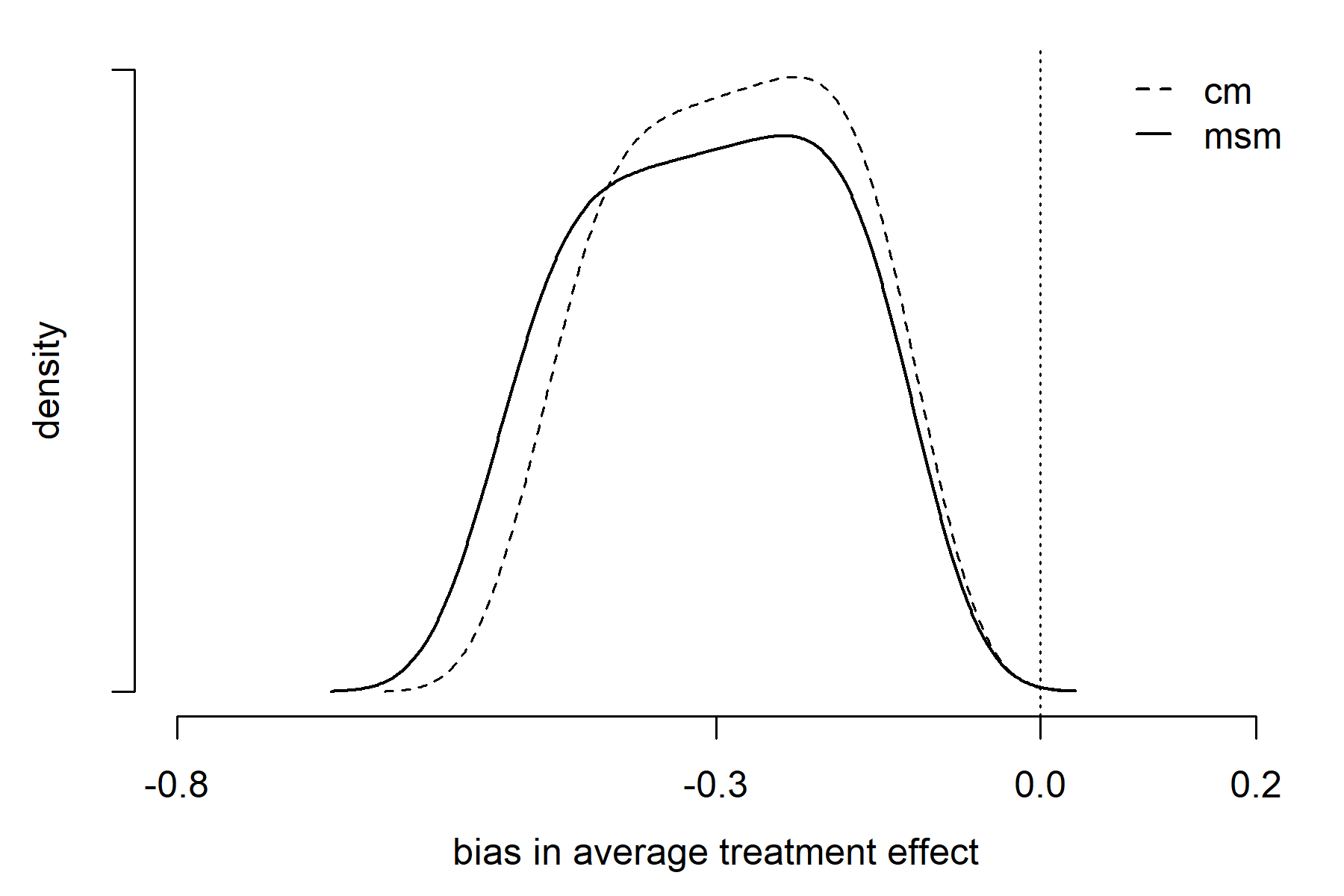}
    \caption{Predicted bias in average treatment effect of diuretics use compared to beta blocker use on mean systolic blood pressure adjusted for self-reported BMI category in NHANES, in a MSM estimated using IPW (msm, solid line) and conditional model (cm, dashed line). The specificity and sensitivity of self-reported BMI category range from $0.90$ to $0.98$.}
    \label{fig:sensmsm}
\end{figure}

By uniformly sampling from the range of possible values of $p_0$ and $p_1$ and applying the bias expressions obtained in section \ref{sec:covme:impact}, a distribution of possible biases is obtained (see appendix section \ref{ap:sec:sensana} for further details). Figure \ref{fig:sensmsm} shows the distribution of bias in a MSM estimated using IPW. Mean bias is -0.31 and median bias is -0.30 (interquartile range -0.40 to -0.20). This result suggests that the results in Table \ref{tab:resultsnhanes} are not affected much by the classification error in self-reported BMI category.

\textbf{BMI measured by trained technicians.} In the NHANES, anthropometric measures were also taken by trained health technicians. By using these measures to calculate BMI category, we found that the specificity of self-reported BMI category was $0.94$ ($p_1$), and the sensitivity was $0.92$ ($p_0 = 0.08$). The ATE (CI) of diuretics use in comparison to beta blocker use on mean blood pressure adjusted for BMI category was -3.59 (-5.84; -1.35) in a MSM estimated using IPW. The ATE was slightly higher than that found by adjusting for self-reported BMI category: -3.52 (-5.74;-1.21), Table \ref{tab:resultsnhanes}. In conclusion, the bias due to classification error was minor, a result in accordance with our sensitivity analysis.

\section{Discussion}
Inverse probability weighting and conditional models are both important and frequently used tools to adjust for confounding variables in observational studies. In this article, we investigated the effect of classification error in a confounding variable in a MSM-IPW in a point-treatment study with a continuous outcome. We derived expressions for the bias in the ATE in a MSM-IPW and a conditional model. These expressions can inform sensitivity analyses for bias due to a misclassified confounding variable.

Sensitivity analysis of misclassified confounding variables is one example of sensitivity analyses for observational epidemiologic studies. Several approaches exist to assess sensitivity of causal conclusions to unmeasured confounding \cite{Groenwold2010SensitivityResearch,VanderWeele2011BiasConfounders,Ding2016SensitivityAssumptions}, that aim to quantify the impact of violations of the assumption of no unmeasured confounding, while our approach aims to quantify the impact of violations of the assumption that all confounding variables are measured without error. Although this paper discusses classification error in a dichotomous confounding variable, the same principles apply to measurement error in a categorical or continuous confounding variable or when multiple confounding variables are considered. Clearly, in such more complex situations, more elaborate assumptions about the structure of measurement error should be made \cite{Keogh2014AEpidemiology}. 

Several methods have been proposed to adjust for measurement error in covariates in MSMs estimated using IPW. Pearl developed a general framework for causal inference in the presence of error-prone covariates, which yields weighted estimators in the case of a dichotomous confounding variable measured with error \cite{Pearl2010OnInference}. The framework relies on a joint distribution of the outcome and the confounding variable. Conversely, the weighting method proposed by McCaffrey et al. does not require a model for the outcome \cite{McCaffrey2013InverseCovariates}. Additionally, regression calibration \cite{Cole2010UsingDeath}, simulation-extrapolation \cite{Kyle2016CorrectingModels,Lockwood2015Simulation-ExtrapolationCovariates} and multiple imputation \cite{Webb-Vargas2017AnAnalysis} have been proposed for correcting for measurement error in covariates of MSMs. These previously discussed methods assume that the measurement error model is known, which may often be unrealistic. In this context it is also important to mention previous studies of the impact of measurement error in the exposure or the endpoint in MSMs, which has been studied by Babanezhad et. al \cite{Babanezhad2010ComparisonMisclassification} and Shu et. al \cite{Shu2017CausalMethods}, respectively.

If treatment is allocated based on an error-prone confounding variable, the treatment effect will not be biased (see DAG in Figure \ref{fig:dags:a}). However, investigators should be careful in concluding that covariate measurement error will not affect their analysis. Suppose that there is an unmeasured variable $U$ that acts as a confounder between the error-prone covariate $L^*$ and treatment $A$. Conditioning on $L^*$ will then open a path between $A$ and $L$ via $U$ and thus confound the relation between $A$ and $Y$. Hence unmeasured variables that would not be problematic if confounding variables were measured without error can introduce unmeasured confounding when confounding variables are measured with error.

This paper considered classification error in a dichotomous confounding variable in a point-treatment study with a continuous outcome. Due to the collapsibility of the marginal treatment effect and conditional treatment effect under the identity link function, the marinal and conditional treatment effect were equal in our study. This is generally not true for models with non-collapsible measures, such as logistic regression. Future research could extend this work to settings with a continuous confounding variable, a binary outcome, multiple confounding variables and to the time-dependent setting with time varying treatments and confounding variables.

The bias expressions derived in this paper can be used to assess bias due to classification error in a dichotomous confounding variable. If classification error in confounding variables is suspected, a sensitivity analysis provides an opportunity to quantitatively inform readers on the possible impact of such errors on causal conclusions.

\newpage
\justify
\begin{singlespacing}
\bibliographystyle{unsrt}
\bibliography{references.bib}

\begin{thebibliography}{10}

\bibitem{Hernan2002EstimatingMeasures}
MA~Hern{\'{a}}n, BA~Brumback, and JM~Robins.
\newblock {Estimating the causal effect of zidovudine on CD4 count with a
  marginal structural model for repeated measures}.
\newblock {\em Stat Med}, 21:1689--1709, 2002.

\bibitem{Robins2000MarginalEpidemiology}
JM~Robins, MA~Hern{\'{a}}n, and BA~Brumback.
\newblock {Marginal structural models and causal inference in Epidemiology}.
\newblock {\em Am J Epidemiol}, 11(5):550--560, 2000.

\bibitem{Rubin2008ForAnalysis}
DB~Rubin.
\newblock {For objective causal inference, design trumps analysis}.
\newblock {\em Ann Appl Stat}, 2(3):808--840, 2008.

\bibitem{Steiner2011OnScores}
PM~Steiner, TD~Cook, and WR~Shadish.
\newblock {On the Importance of reliable covariate measurement in selection
  bias adjustments using propensity scores}.
\newblock {\em J Educ Behav Stat}, 36(2):213--236, 2011.

\bibitem{Michels2001AError}
KB~Michels.
\newblock {A renaissance for measurement error}.
\newblock {\em Int J Epidemiol}, 30(3):421--422, 2001.

\bibitem{vanSmeden2019ReflectionResearch}
M~van Smeden, TL~Lash, and RHH Groenwold.
\newblock {Reflection on modern methods: Five myths about measurement error in
  epidemiologic research}.
\newblock {\em Int J Epidemiol}, forthcoming, 2019.

\bibitem{Raboud1995QuantificationIndividuals}
JM~Raboud, L~Haley, JS~Montaner, C~Murphy, M~Januszewska, and MT~Schechter.
\newblock {Quantification of the variation due to laboratory and physiologic
  sources in CD4 lymphocyte counts of clinically stable HIV-infected
  individuals}.
\newblock {\em J Acquir Immune Defic Syndr Hum Retrovirol}, 10(Suppl 2):67--73,
  1995.

\bibitem{Kyle2016CorrectingModels}
RP~Kyle, EEM Moodie, MB~Klein, and M~Abrahamowicz.
\newblock {Correcting for measurement error in time-varying covariates in
  marginal structural models}.
\newblock {\em Am J Epidemiol}, 184(3):249--258, 2016.

\bibitem{Buonaccorsi2010MeasurementApplications}
JP~Buonaccorsi.
\newblock {\em {Measurement error: Models, methods, and applications}}.
\newblock Chapman {\&} Hall/CRC, Boca Raton, FL, 2010.

\bibitem{Carroll2006}
RJ~Carroll, D~Ruppert, LA~Stefanski, and CM~Crainiceanu.
\newblock {\em {Measurement error in nonlinear models: A modern perspective}}.
\newblock Chapman {\&} Hall/CRC, Boca Raton, FL, 2nd edition, 2006.

\bibitem{Gustafson2004MeasurementAdjustments.}
P~Gustafson.
\newblock {\em {Measurement error and misclassification in statistics and
  epidemiology: Impacts and Bayesian adjustments.}}
\newblock Chapman {\&} Hall/CRC, Boca Raton, FL, 2004.

\bibitem{Fuller1987MeasurementModels}
WA~Fuller.
\newblock {\em {Measurement Error Models}}.
\newblock John Wiley {\&} Sons, New York, NY, 1987.

\bibitem{Regier2014TheStudy}
MD~Regier, EEM Moodie, and RW~Platt.
\newblock {The effect of error-in-confounders on the estimation of the causal
  parameter when using marginal structural models and inverse probability-of-
  treatment weights: A simulation study}.
\newblock {\em Int J Biostat}, 10(1):1--15, 2014.

\bibitem{McCaffrey2013InverseCovariates}
DF~McCaffrey, JR~Lockwood, and CM~Setodji.
\newblock {Inverse probability weighting with error-prone covariates}.
\newblock {\em Biometrika}, 100(3):671--680, 2013.

\bibitem{Greenland1996BasicBiases}
S~Greenland.
\newblock {Basic methods for sensitivity analysis of biases}.
\newblock {\em Int J Epidemiol}, 25(6):1107--1116, 1996.

\bibitem{Lash2009ApplyingData}
TL~Lash, MP~Fox, and AK~Fink.
\newblock {\em {Applying quantitative bias analysis to epidemiologic data}}.
\newblock Springer, New York, NY, 2009.

\bibitem{Lash2014GoodAnalysis}
TL~Lash, MP~Fox, RF~Maclehose, G~Maldonado, LC~Mccandless, and S~Greenland.
\newblock {Good practices for quantitative bias analysis}.
\newblock {\em Int J Epidemiol}, 43(6):1969--1985, 2014.

\bibitem{Jackson2006EvidenceSeniors}
LA~Jackson, ML~Jackson, JC~Nelson, KM~Neuzil, and NS~Weiss.
\newblock {Evidence of bias in estimates of influenza vaccine effectiveness in
  seniors}.
\newblock {\em Int J Epidemiol}, 35(2):337--344, 4 2006.

\bibitem{Rubin1974EstimatingStudies}
DB~Rubin.
\newblock {Estimating causal effects of treatments in randomized and
  nonrandomized studies}.
\newblock {\em J Educ Psychol}, 66(5):688--701, 1974.

\bibitem{Rubin1990FormalEffects}
DB~Rubin.
\newblock {Formal modes of statistical inference for causal effects}.
\newblock {\em J Stat Plan Inference}, 25:279--292, 1990.

\bibitem{Hernan2019CausalInference}
MA~Hern{\aa}n and JM~Robins.
\newblock {\em {Causal Inference}}.
\newblock Chapman {\&} Hall/CRC, Boca Raton, forthcoming edition, 2019.

\bibitem{Rubin1980RandomizationComment}
DB~Rubin.
\newblock {Randomization analysis of experimental data: The Fisher
  randomization test comment}.
\newblock {\em J Am Stat Assoc}, 75(371):591--593, 1980.

\bibitem{Cole2008ConstructingModels}
SR~Cole and MA~Hern{\'{a}}n.
\newblock {Constructing inverse probability weights for marginal structural
  models}.
\newblock {\em Am J Epidemiol}, 168(6):656--664, 9 2008.

\bibitem{Steiner2016TheBiases}
PM~Steiner and Y~Kim.
\newblock {The mechanics of omitted variable bias: bias amplification and
  cancellation of offsetting biases}.
\newblock {\em J Causal Inference}, 4(2):20160009, 2016.

\bibitem{RCoreTeam2018R:Computing}
{R Core Team}.
\newblock {R: A language and environment for statistical computing}, 2018.

\bibitem{VanDerWal2011Ipw:Weighting}
WM~Van Der~Wal and RB~Geskus.
\newblock {ipw: An R package for inverse probability weighting}.
\newblock {\em J Stat Softw}, 43(13):1--23, 2011.

\bibitem{Lumley2004AnalysisSamples}
T~Lumley.
\newblock {Analysis of complex survey samples}.
\newblock {\em J Stat Softw}, 9(1):1--19, 2004.

\bibitem{Morris2019UsingMethods}
TP~Morris, IR~White, and MJ~Crowther.
\newblock {Using simulation studies to evaluate statistical methods}.
\newblock {\em Stat Med}, pages 1--29, 2019.

\bibitem{Gasparini2018Rsimsum:Studies}
A~Gasparini.
\newblock {rsimsum: Summarise results from Monte Carlo simulation studies}.
\newblock {\em JOSS}, 3(26):739, 6 2018.

\bibitem{Robins2000MarginalInference}
JM~Robins.
\newblock {Marginal structural models versus structural nested models as tools
  for causal inference}.
\newblock In EM~Halloran and D~Berry, editors, {\em Statistical models in
  epidemiology, the environment, and clinical trials}, chapter~2, pages
  95--133. Springer-Verlag, New York, 1 edition, 2000.

\bibitem{Centershttps://wwwn.cdc.gov/nchs/nhanes/continuousnhanes/default.aspxBeginYear=2011.}
{Centers for Disease Control and Prevention (CDC). National Center for Health
  Statistics (NCHS). National Health and Nutrition Examination Survey Data.
  Hyattsville, MD: U.S. Department of Health and Human Services, Centers for
  Disease Control and Prevention, [2011-2012],
  [https://wwwn.cdc.gov/nchs/nhanes/continuousnhanes/default.aspx?BeginYear=2011].}

\bibitem{Centershttps://wwwn.cdc.gov/nchs/nhanes/continuousnhanes/default.aspxBeginYear=2013.}
{Centers for Disease Control and Prevention (CDC). National Center for Health
  Statistics (NCHS). National Health and Nutrition Examination Survey Data.
  Hyattsville, MD: U.S. Department of Health and Human Services, Centers for
  Disease Control and Prevention, [2013-2014],
  [https://wwwn.cdc.gov/nchs/nhanes/continuousnhanes/default.aspx?BeginYear=2013].}

\bibitem{Altman2006TheVariables}
DG~Altman and P~Royston.
\newblock {The cost of dichotomising continuous variables}.
\newblock {\em BMJ}, 332:1080, 2006.

\bibitem{Groenwold2010SensitivityResearch}
RHH Groenwold, DB~Nelson, KL~Nichol, AW~Hoes, and E~Hak.
\newblock {Sensitivity analyses to estimate the potential impact of unmeasured
  confounding in causal research}.
\newblock {\em Int J Epidemiol}, 39(1):107--117, 2 2010.

\bibitem{VanderWeele2011BiasConfounders}
TJ~VanderWeele and OA~Arah.
\newblock {Bias formulas for sensitivity analysis of unmeasured confounding for
  general outcomes, treatments, and confounders}.
\newblock {\em Epidemiology}, 22(1):42--52, 2011.

\bibitem{Ding2016SensitivityAssumptions}
P~Ding and TJ~VanderWeele.
\newblock {Sensitivity analysis without assumptions}.
\newblock {\em Epidemiology}, 27(3):368--377, 2016.

\bibitem{Keogh2014AEpidemiology}
RH~Keogh and IR~White.
\newblock {A toolkit for measurement error correction, with a focus on
  nutritional epidemiology}.
\newblock {\em Stat Med}, 33:2137--2155, 2014.

\bibitem{Pearl2010OnInference}
J~Pearl.
\newblock {On measurement bias in causal inference}.
\newblock Technical report, AUAI, Corvallis, OR, 2010.

\bibitem{Cole2010UsingDeath}
SR~Cole, LP~Jacobson, PC~Tien, L~Kingsley, JS~Chmiel, and K~Anastos.
\newblock {Using marginal structural measurement-error models to estimate the
  long-term effect of antiretroviral therapy on incident AIDS or death}.
\newblock {\em Am J Epidemiol}, 171(1):113--122, 2010.

\bibitem{Lockwood2015Simulation-ExtrapolationCovariates}
JR~Lockwood and DF~McCaffrey.
\newblock {Simulation-Extrapolation for estimating means and causal effects
  with mismeasured covariates}.
\newblock {\em Obs Stud}, 1:241--290, 2015.

\bibitem{Webb-Vargas2017AnAnalysis}
Y~Webb-Vargas, KE~Rudolph, D~Lenis, P~Murakami, and EA~Stuart.
\newblock {An imputation-based solution to using mismeasured covariates in
  propensity score analysis}.
\newblock {\em Stat Methods Med Res}, 26(4):1824--1837, 2017.

\bibitem{Babanezhad2010ComparisonMisclassification}
M~Babanezhad, S~Vansteelandt, and E~Goetghebeur.
\newblock {Comparison of causal effect estimators under exposure
  misclassification}.
\newblock {\em J Stat Plan Inference}, 140:1306--1319, 2010.

\bibitem{Shu2017CausalMethods}
Di~Shu and Grace~Y. Yi.
\newblock {Causal inference with measurement error in outcomes: Bias analysis
  and estimation methods}, 2017.

\end{thebibliography}
\begin{table}[hb]
    \centering
    \caption{Values of the parameters in the five different simulation scenarios}
    \begin{tabular}{clllllll}
         \textbf{Scenario} & $p_0$ & $p_1$ & $\lambda$ & $\pi_0$ & $\pi_1$ & $\beta$ & $\gamma$ \\
         \hline
         \textbf{0} & 0 & 1 & 0.50 & 0.50 & 0.75 & 1 & 2\\
         \textbf{1} & 0.05 & 0.90 & 0.50 & 0.90 & 0.45 & 1 & 2 \\
         \textbf{2} & 0.05 & 0.90 & 0.80 & 0.25 & 0.75 & 1 & 2 \\
         \textbf{3} & 0.05 & 0.90 & 0.80 & 0.50 & 0.75 & 1 & 2 \\
         \textbf{4} & 0.05 & 0.90 & 0.45 & 0.50 & 0.75 & 1 & 2\\
    \end{tabular}
    \label{tab:simscen}
\end{table}

\afterpage{%
\begin{landscape}
\centering
\begin{singlespacing}
\begin{threeparttable}[p]
\caption{Results of simulation study studying the performance of MSMs estimated using IPW (MSM) and conditional models (CM) if there is classification error in the confounding variable, estimated generating 5,000 data sets with  different sample sizes (n = 1,000 and n = 100). Bias (formula) = bias based on bias expressions derived in section \ref{sec:covme:impact}; MSE = mean squared error; empSE = empirical standard error; modelSE = model based standard error. Monte Carlo standard errors are shown between brackets. The parameters in scenarios 0-4 are explained in the footnote of the table.}\label{tab:ressim}
\centering
\begin{tabular}{llrrrrrrr}
\textbf{Method} & \textbf{Sample size} & \textbf{Scenario}\tnote{a} & \textbf{Bias (formula)} & \textbf{Bias} & \textbf{MSE} & \textbf{Coverage} & \textbf{empSE} & \textbf{modelSE} \\ 
  \hline
\rowcolor{lightgray}
\textbf{MSM} & 1,000 & \textbf{0} & 0.00 & 0.00 (0.001) & 0.00 (0.000) & 0.99 (0.001) & 0.07 (0.001) & 0.10 (0.000) \\ 
\rowcolor{lightgray}
 & & \textbf{1} & -0.42 & -0.42 (0.001) & 0.18 (0.001) & 0.03 (0.002) & 0.10 (0.001) & 0.11 (0.000) \\ 
\rowcolor{lightgray}
 & & \textbf{2} & 0.14 & 0.14 (0.001) & 0.03 (0.000) & 0.67 (0.007) & 0.08 (0.001) & 0.09 (0.000) \\ 
\rowcolor{lightgray}
 & & \textbf{3} & 0.29 & 0.29 (0.001) & 0.09 (0.001) & 0.08 (0.004) & 0.08 (0.001) & 0.09 (0.000) \\ 
\rowcolor{lightgray}
 & & \textbf{4} & 0.15 & 0.15 (0.001) & 0.03 (0.000) & 0.68 (0.007) & 0.08 (0.001) & 0.10 (0.000) \\ 
 & 100 & \textbf{0} & 0.00 & 0.00 (0.003) & 0.05 (0.001) & 0.99 (0.001) & 0.22 (0.002) & 0.31 (0.000) \\ 
 & & \textbf{1} & -0.42 & -0.42 (0.005) & 0.29 (0.005) & 0.78 (0.006) & 0.34 (0.003) & 0.37 (0.001) \\ 
 & & \textbf{2} & 0.14 & 0.14 (0.004) & 0.08 (0.002) & 0.94 (0.003) & 0.25 (0.003) & 0.29 (0.000) \\ 
 & & \textbf{3} & 0.29 & 0.29 (0.004) & 0.15 (0.002) & 0.84 (0.005) & 0.26 (0.003) & 0.28 (0.000) \\ 
 & & \textbf{4} & 0.15 & 0.15 (0.004) & 0.08 (0.002) & 0.95 (0.003) & 0.25 (0.002) & 0.31 (0.000) \\ 
\rowcolor{lightgray}
\textbf{CM} & 1,000 & \textbf{0} & 0.00 & 0.00 (0.001) & 0.00 (0.000) & 0.95 (0.003) & 0.07 (0.001) & 0.07 (0.000) \\ 
\rowcolor{lightgray}
 & & \textbf{1} & -0.34 & -0.34 (0.001) & 0.12 (0.001) & 0.02 (0.002) & 0.09 (0.001) & 0.08 (0.000) \\ 
\rowcolor{lightgray}
 & & \textbf{2} & 0.16 & 0.16 (0.001) & 0.03 (0.000) & 0.46 (0.007) & 0.08 (0.001) & 0.08 (0.000) \\ 
\rowcolor{lightgray}
 & & \textbf{3} & 0.32 & 0.32 (0.001) & 0.11 (0.001) & 0.02 (0.002) & 0.08 (0.001) & 0.08 (0.000) \\ 
\rowcolor{lightgray}
 & & \textbf{4} & 0.15 & 0.15 (0.001) & 0.03 (0.000) & 0.49 (0.007) & 0.08 (0.001) & 0.07 (0.000) \\ 
 & 100 & \textbf{0} & 0.00 & 0.00 (0.003) & 0.05 (0.001) & 0.95 (0.003) & 0.22 (0.002) & 0.22 (0.000) \\ 
 & & \textbf{1} & -0.34 & -0.33 (0.004) & 0.19 (0.003) & 0.73 (0.006) & 0.29 (0.003) & 0.27 (0.000) \\ 
 & & \textbf{2} & 0.16 & 0.16 (0.004) & 0.09 (0.002) & 0.90 (0.004) & 0.25 (0.003) & 0.25 (0.000) \\ 
 & & \textbf{3} & 0.32 & 0.32 (0.004) & 0.17 (0.003) & 0.74 (0.006) & 0.26 (0.003) & 0.25 (0.000) \\ 
 & & \textbf{4} & 0.15 & 0.15 (0.003) & 0.08 (0.002) & 0.90 (0.004) & 0.24 (0.002) & 0.24 (0.000) \\ 
   \hline
\end{tabular}
  \begin{tablenotes}
\scriptsize
    \item[a] In all scenarios, the average treatment effect (estimand) is 1 ($\beta = 1$) and the effect of the confounding variable on the outcome is 2 ($\gamma =2$). In scenario 0, there is no classification error (specificity and sensitivity of the misclassified confounding variable are 1, i.e., $p_0 = 0$ and $p_1 =1$). In scenarios 1-4, the specificity of the misclassified confounding variable is 0.95 (i.e., $p_0 = 0.05$) and the sensitivity is 0.9 (i.e., $p_1 = 0.9$). The prevalence of the confounding variable ($\lambda$), and the probability of receiving treatment if the confounding is not present or present ($\pi_0$ and $\pi_1$, respectively) are set as follows in the scenarios: scenario 0: $\lambda = 0.5$, $\pi_0 = 0.5$, $\pi_1 = 0.75$; scenario 1: $\lambda = 0.5$, $\pi_0 = 0.9$, $\pi_1 = 0.45$; scenario 2: $\lambda = 0.8$, $\pi_0 = 0.25$, $\pi_1 = 0.75$; scenario 3: $\lambda = 0.8$, $\pi_0 = 0.5$, $\pi_1 = 0.75$; scenario 4: $\lambda = 0.45$, $\pi_0 = 0.5$, $\pi_1 = 0.75$.
  \end{tablenotes}
\end{threeparttable}
\end{singlespacing}
\end{landscape}
}
\begin{table}[ht]
\centering
\begin{singlespacing}
\begin{threeparttable}
\caption{Average treatment effect of diuretics use compared to beta blocker use on mean systolic blood pressure unadjusted and adjusted for self-reported categorised BMI}\label{tab:resultsnhanes}
    \begin{tabular}{ll}
         \textbf{Model} & \textbf{Effect size (CI)} \\
         \hline
         Unadjusted & -4.03 (-6.30; -1.76) \\
         Marginal structural model\tnote{a} & -3.52 (-1.21; -5.74)\\
         Conditional model & -3.48 (-1.27; -5.76)
    \end{tabular}
    \label{tab:effects}
  \begin{tablenotes}
\scriptsize
    \item[a] estimated using inverse probability weighting
  \end{tablenotes}
\end{threeparttable}
\end{singlespacing}
\end{table}

\end{singlespacing}

\newpage
\appendix
\renewcommand{\thesection}{A\arabic{section}}
\renewcommand{\thesubsection}{A\arabic{section}.\arabic{subsection}}
\renewcommand{\theequation}{A\arabic{equation}}
\setcounter{equation}{0}
\begin{singlespacing}
\section{Quantification of bias due to classification error in a confounding
variable}\label{appendix}
\end{singlespacing}

\subsection{Conditional model}\label{ap:sec:cm}
Under the assumptions in section \ref{sec:covme} and by the law of total expectation, the expected value of the outcome $Y$ given the covariates $A$ and $L^*$ is,
\begin{eqnarray*}
    \E[Y|A=a, L^*] = \E_{L|A=a, L^*}[\E[Y|A=a, L^*, L]]
    &=&\E_{L|A=a, L^*}[\alpha + \beta a + \gamma L]\nonumber\\
    &=&\alpha + \beta a + \gamma\E[L|A=a, L^*]\\
    &=&\alpha + \beta a + \gamma\phi_{aL^*}\\
    &=&\{\alpha + \gamma \phi_{00}\} + \{\beta + \gamma(\phi_{10} - \phi_{00})\}a \\
    &+& \{\gamma(\phi_{01}-\phi_{00})\}L^*+ \gamma(\phi_{11}-\phi_{10}-\phi_{01}+\phi_{00})aL^*,
\end{eqnarray*}
which relies on the assumption that $L^*$ is non-differentially misclassified with respect to the outcome and includes an interaction between $A$ and $L^*$. Further, $\phi_{al^*}$ is the probability that confounding variable $L$ is one, given that treatment $A$ is $a$ and that misclassfied confounding variable $L^*$ is $l^*$, or,
\begin{eqnarray*}\label{apeq:phi_als}
\phi_{al^*} &=& P(L = 1|A=a, L^*=l^*) \nonumber\\
&=& \frac{P(A=a|L=1, L^*=l^*)P(L=1|L^*=l^*)}{P(A=a|L^*=l^*)}\nonumber\\
&=& \frac{P(A=a|L=1)P(L=1|L^*=l^*)}{P(A=a|L^*=l^*)}\nonumber\\
&=& \frac{P(A=a|L=1)\frac{P(L^*=l^*|L=1)P(L=1)}{P(L^*=l^*)}}{P(A=a|L^*=l^*)}\nonumber\\
&=& \frac{P(A=a|L=1)P(L^*=l^*|L=1)P(L=1)}{P(A=a|L^*=l^*)P(L^*=l^*)}\nonumber\\
&=& \frac{\lambda(1-\pi_1)^{(1-a)}\pi_1^{a} (1-p_1)^{(1-l^*)} p_1^{l^*}}{(1-\pi^*_{l^*})^{(1-a)}{\pi^*_{l^*}}^a(1-\ell)^{(1-l^*)}\ell^{l^*}}.
\end{eqnarray*}
Where $\ell=P(L^*=l^*)=p_0(1-\lambda)+p_1\lambda$ and $\pi^*_{l^*}$ is the probability that treatment $A$ is one given that the misclassified confounding variable $L^*=l^*$ or, 
\begin{eqnarray*}
\pi^*_{l^*} &=& P(A=1|L^*=l^*) \nonumber \\ 
&=&\Sigma_lP(A=1|L^*=l^*,L=l)P(L=l|L^*=l^*) \nonumber \\
&=&\Sigma_lP(A=1|L=l)P(L=l|L^*=l^*) \nonumber \\
&=&\Sigma_lP(A=1|L=l) \frac{P(L^*=l^*|L=l)P(L=l)}{P(L^*=l^*)} \nonumber \\
&=&\Sigma_l\pi_l \frac{(1-p_l)^{(1-l^*)}p_l^{l^*}(1-\lambda)^{1-l}\lambda^l}{(1-\ell)^{1-l^*}\ell^{l^*}}.
\end{eqnarray*}
Thus, the bias in the coefficient for $A=a$ is $\gamma(\phi_{10} - \phi_{00})$ if one includes an interaction term between $A$ and $L^*$ to model mean $Y$ given $A$ and $L^*$. Yet, in this model, the coefficient for $A=a$ now represents the ATE given that $L^*$ is null. Typically, one would only include main effects of $A$ and $L^*$ in a conditional model for $Y$ conditional on $A$ and $L^*$:
\begin{eqnarray*}\label{appeq:melm}
    \E[Y|A=a, L^*] &=& \{\alpha + \gamma \phi_{00}\} + \{\beta + \gamma(\phi_{10} - \phi_{00})\}a
    + \{\gamma(\phi_{01}-\phi_{00})\}L^*\\
    &+& \gamma(\phi_{11}-\phi_{10}-\phi_{01}+\phi_{00})aL^*\\
    &=& \{\alpha + \gamma \phi_{00}\} + \{\beta + \gamma(\phi_{10} - \phi_{00})\}a
    + \{\gamma(\phi_{01}-\phi_{00})\}L^*\\
    &+& \gamma(\phi_{11}-\phi_{10}-\phi_{01}+\phi_{00})\E[aL^*|A=a, L]\\
    &=&\{\alpha + \gamma \phi_{00} + \delta u_0\}
    + \{\beta + \gamma (\phi_{10} - \phi_{00}) + \delta u_A\}a\\
    &+& \{\gamma(\phi_{01}-\phi_{00}) + \delta u_{L^*}\}L^*,
\end{eqnarray*}
where $u_0, u_A, u_{L^*}$ are the coefficients of the linear model $\E[AL^*|AL]=u_0 + u_AA + u_{L^*}L$ and $\delta = \gamma(\phi_{11}-\phi_{10}-\phi_{01}+\phi_{00})$. Here, 
\begin{eqnarray*}
u_A &=& \frac{\Var(L^*)\Cov(A,AL^*)-\Cov(A,L^*)\Cov(L^*,AL^*)}{\Var(L^*)\Var(A)-\Cov(A,L^*)^2}, \\
u_{L^*} &=& \frac{\Var(A)\Cov(L^*,AL^*)-\Cov(A,L^*)\Cov(A,AL^*)}{\Var(L^*)\Var(A)-\Cov(A,L^*)^2}, \\
u_0 &=& \overline{AL^*}-u_A\overline{A}-u_{L^*}\overline{L^*},
\end{eqnarray*}
where $\overline{AL^*}, \overline{A}$ and $\overline{L^*}$ denote mean of $A$ times $L^*$, $A$ and $L^*$, respectively. Now, we like to express $u_A$ and $u_{L^*}$ in terms of $\lambda$, $\pi_0$, $\pi_1$, $p_0$ and $p_1$. We can write a linear model for the model of $A$ conditional on $L^*$ using that $P(A=1|L^*=l^*)=\pi^*_{l^*}$ and use standard regression theory to get an expression for $\Cov(A,L^*)$:
\begin{eqnarray*}
\E[A|L^*] = \pi^*_0 + (\pi^*_1 - \pi^*_0)L^*, \quad
\pi^*_1 - \pi^*_0 = \frac{\Cov(A,L^*)}{\Var(L^*)}, \quad \text{thus} \quad
\Cov(A,L^*) = (\pi^*_1 - \pi^*_0)\Var(L^*).
\end{eqnarray*}
Where, $\Var(L^*) = \ell (1-\ell)$. Since $\E[AL^*|L=0]=0$ and $\E[AL^*|L=1]=\E[A|L^*=1]=\pi^*_1$, it follows that,
\begin{eqnarray*}
\E[AL^*|L^*] &=& \pi^*_1 L^*, \quad \pi^*_1 = \frac{\Cov(AL^*,L^*)}{\Var(L^*)}, \quad \text{thus} \quad
\Cov(AL^*,L^*) = \pi^*_1 \Var(L^*).
\end{eqnarray*}
Equivalently, since $\E[AL^*|A=0]=0$ and $\E[AL^*|A=1]=\E[L^*|A=1]$, it follows that,
\begin{eqnarray*}
\E[AL^*|A] &=& \E[L^*|A=1] A = \frac{P(A=1|L^*=1)P(L^*=1)}{P(A=1)} A,\\
\E[L^*|A=1] &=& \frac{\pi^*_1 \ell}{\omega}, \quad
\frac{\pi^*_1 \ell}{a} = \frac{\Cov(AL^*,A)}{\Var(A)}, \quad \text{thus} \quad
\Cov(AL^*,A) = \frac{\pi^*_1 \ell}{\omega}\Var(A).
\end{eqnarray*}
Where, $\Var(A) = \omega (1-\omega)$, and $\omega=P(A=1)=\pi_0(1-\lambda)+\pi_1\lambda$. To conclude, 
\begin{eqnarray*}
u_A &=& \frac{\pi_1^*\ell/\omega \Var(A)\Var(L^*) -\pi_1^*(\pi^*_1-\pi^*_0)\Var(L^*)^2}{\Var(A)\Var(L^*)-(\pi_1 - \pi_0)^2 \Var(L^*)^2}\\
&=& \frac{\pi_1^*\ell/\omega \Var(A) -\pi_1^*(\pi^*_1-\pi^*_0)\Var(L^*)}{\Var(A)-(\pi_1 - \pi_0)^2 \Var(L^*)}\\
&=&\frac{\pi^*_1\ell(1-\omega)-\pi^*_1(\pi^*_1-\pi^*_0)\ell(1-\ell))}{\omega(1-\omega) - (\pi^*_1-\pi^*_0)^2\ell(1-\ell) } \\
u_{L^*} &=& \frac{\pi^*_1\Var(A)\Var(L^*) -\pi^*_1\ell/\omega(\pi^*_1-\pi^*_0)\Var(A)\Var(L^*)}{\Var(L^*)\Var(A)-((\pi^*_1-\pi^*_0)\Var(L^*))^2}, \\
u_0 &=& \overline{AL^*}-u_A\overline{A}-u_{L^*}\overline{L^*}.
\end{eqnarray*}
The intercept, the coefficient for $A$ and the coefficient for $L^*$ in a conditional model for $Y$ conditional on $A$ and $L^*$ which includes only main effects of $A$ and $L^*$ are, respectively:
\begin{eqnarray*}
\alpha + \gamma\phi_{00} + \delta u_0,\\
\beta + \gamma(\phi_{10}-\phi_{00})(1- \frac{\pi^*_1\ell(1-\omega)-\pi^*_1(\pi^*_1-\pi^*_0)\ell(1-\ell))}{\omega(1-\omega) - (\pi^*_1-\pi^*_0)^2\ell(1-\ell) })\\
+ \gamma(\phi_{11}-\phi_{01})(\frac{\pi^*_1\ell(1-\omega)-\pi^*_1(\pi^*_1-\pi^*_0)\ell(1-\ell))}{\omega(1-\omega) - (\pi^*_1-\pi^*_0)^2\ell(1-\ell) }),\\
\text{and} \quad \gamma(\phi_{01} - \phi_{00}) + \delta u_{L^*}.
\end{eqnarray*}

\subsection{MSM}\label{ap:sec:msm}
Under the assumptions in section \ref{sec:covme}, a MSM under model (\ref{eq:msm}) proceeds by solving the weighted regression model, 
\begin{eqnarray*}
    \sum_{i=1}^n \frac{1}{P(A_i|L_i^*)}(Y_i - \alpha_{\text{msm}} - \beta A_i) = 0 \quad \text{and} \quad
    \sum_{i=1}^n \frac{A_i}{P(A_i|L_i^*)}(Y_i - \alpha_{\text{msm}} - \beta A_i) = 0.
\end{eqnarray*}
Solving these equations for $\alpha_{\text{msm}}$ and $\beta$ result in the following estimators:
\begin{eqnarray*}
    \hat{\alpha}_{\mathrm{msm}}=\overline{Y}_{w^*}-\hat{\beta}_{\mathrm{msm}}\overline{A}_{w^*} \quad \text{and} \quad
    \hat{\beta}=\frac{\sum_{i=1}^{n}\frac{1}{P(A_i|L_i)}(Y_i-\overline{Y}_{w^*})(A_i-\overline{A}_{w^*})}{\sum_{i=1}^{n}\frac{1}{P(A_i|L_i)}(A_i-\overline{A}_{w^*})^2}.
\end{eqnarray*}
Where, 
\begin{eqnarray*}
    \overline{Y}_{w^*}=\frac{\sum_{i=1}^{n}Y_i/P(A_i|L_i^*)}{\sum_{i=1}^{n}1/P(A_i|L_i^*)} \quad \text{and} \quad 
    \overline{A}_{w^*}=\frac{\sum_{i=1}^{n}A_i/P(A_i|L_i^*)}{\sum_{i=1}^{n}1/P(A_i|L_i^*)}.
\end{eqnarray*}
Let $n^*_{al}$ be the number of people with $A=a$ and $L^*=l^*$. In a population of $n$ people, 
\begin{eqnarray*}
n^*_{00} &=& n P(A=0, L^*=0) = n P(A=0| L^*=0)P(L^*=0)\\
&=& n \sum^l P(A=0|L=l, L^*=0)P(L=l|L^*=0)\\
&=& n \sum^l P(A=0|L=l)P(L=l|L^*=0)P(L^*=0)\\
&=& n \sum^l P(A=0|L=l)P(L=l)P(L^*=0|L=l)\\
&=& n_{00}(1-p_0) + n_{01}(1-p_1),\\
n^*_{01} &=& n_{00}p_0 + n_{01} p_1, \quad
n^*_{10} = n_{10}(1-p_0) + n_{11}(1-p_1) \quad \text{and} \quad
n^*_{11} = n_{10}p_0 + n_{11} p_1.
\end{eqnarray*}
Hence, 
\begin{eqnarray*}
\sum_{i=1}^{n}1/P(A_i|L_i^*) &=& \sum_{i=1}^{n}\frac{1}{\sum_l[P(A_i|L_i^*,L=l)P(L=l|L_i^*)]}\\
&=& \sum_{i=1}^{n}\frac{1}{\sum_l[P(A_i|L=l)P(L=l|L_i^*)]}\\
&=& \sum^{n^*_{00}} \frac{1}{\sum_l[(1-\pi_l)P(L=l|L^*=0)]}+\sum^{n^*_{01}} \frac{1}{\sum_l[(1-\pi_l)P(L=l|L^*=1)]}\\
&+& \sum^{n^*_{10}} \frac{1}{\sum_l[\pi_lP(L=l|L^*=0)]}+\sum^{n^*_{11}} \frac{1}{\sum_l[\pi_lP(L=l|L^*=1)]}.
\end{eqnarray*}
Where, 
\begin{eqnarray*}
\sum^{n^*_{00}} \frac{1}{\sum_l[(1-\pi_l)P(L=l|L^*=0)]}&=&\frac{n_{00}(1-p_0)+n_{01}(1-p_1)}{(1-\pi_0)P(L=0|L^*=0) + (1-\pi_1)P(L=1|L^*=0)}=\\
&=&\frac{n_{00}(1-p_0)+n_{01}(1-p_1)}{(1-\pi_0)\frac{P(L^*=0|L=0)(1-\lambda)}{P(L^*=0)} + (1-\pi_1)\frac{P(L^*=0|L=1)\lambda}{P(L^*=0)}}=\\
&=&\frac{n_{00}(1-p_0)+n_{01}(1-p_1)}{\frac{n_{00}}{nP(L^*=0)}(1-p_0) + \frac{n_{01}}{nP(L^*=0)}(1-p_1)}\\
&=&\frac{1}{1/(nP(L^*=0))}\\
&=&nP(L^*=0)=n(1-\ell),\\
\sum^{n^*_{01}} \frac{1}{\sum_l[(1-\pi_l)P(L=l|L^*=1)]}&=&nP(L^*=1)=n\ell,\\
\sum^{n^*_{10}} \frac{1}{\sum_l[\pi_lP(L=l|L^*=0)]}&=&nP(L^*=0)=n(1-\ell), \\
\sum^{n^*_{11}} \frac{1}{\sum_l[\pi_lP(L=l|L^*=1)]}&=&nP(L^*=1)=n\ell.\\
\end{eqnarray*}
Frow which it follows that, 
\begin{eqnarray*}
\sum_{i=1}^{n}1/P(A_i|L_i^*) = 2n(1-\ell)+2n\ell=2n.
\end{eqnarray*}
Further, 
\begin{eqnarray*}
\sum_{i=1}^{n}\E[Y_i]/P(A_i|L^*_i) &=& \sum^{n^*_{00}} \frac{\E[Y_i]}{\sum_l[(1-\pi_l)P(L=l|L^*=0)]}+\sum^{n^*_{01}} \frac{\E[Y_i]}{\sum_l[(1-\pi_l)P(L=l|L^*=1)]}\\
&+& \sum^{n^*_{10}} \frac{\E[Y_i]}{\sum_l[\pi_lP(L=l|L^*=0)]}+\sum^{n^*_{11}} \frac{\E[Y_i]}{\sum_l[\pi_lP(L=l|L^*=1)]}\\
&=& \sum^{n^*_{00}} \frac{\alpha + \gamma P(L=1|A=0, L^*=0)}{\sum_l[(1-\pi_l)P(L=l|L^*=0)]}+\sum^{n^*_{01}} \frac{\alpha + \gamma P(L=1|A=0, L^*=1)}{\sum_l[(1-\pi_l)P(L=l|L^*=1)]}\\
&+& \sum^{n^*_{10}} \frac{\alpha + \beta + \gamma P(L=1|A=1, L^*=0)}{\sum_l[\pi_lP(L=l|L^*=0)]}+\sum^{n^*_{11}} \frac{\alpha + \beta + \gamma P(L=1| A=1, L^*=1)}{\sum_l[\pi_lP(L=l|L^*=1)]}\\
&=& n\alpha (1-\ell) + n\gamma (1-\ell)\phi_{00}
+ n\alpha \ell + n\gamma \phi_{01} 
+ n(\alpha + \beta)(1-\ell) + n\gamma (1-\ell)\phi_{10}\\ 
&+& n(\alpha + \beta)\ell + n\gamma \phi_{11} \\
&=& 2n\alpha + n\beta + n\gamma(1-\ell)(\phi_{00}
+\phi_{10}) + n\gamma\ell(\phi_{01}
+ \phi_{11})
\end{eqnarray*}
And,
\begin{eqnarray*}
\sum_{i=1}^{n}A_i/P(A_i|L_i) &=& \sum^{n^*_{10}} \frac{1}{\sum_l[\pi_lP(L=l|L^*=0)]}+\sum^{n^*_{11}} \frac{1}{\sum_l[\pi_lP(L=l|L^*=1)]}\\
&=& n(1-p_0)(1-\lambda)+n(1-p_1)\lambda + np_0(1-\lambda)+np_1\lambda = n.
\end{eqnarray*}
Thus,
\begin{eqnarray*}
\E[\overline{Y}_{w^*}]= \alpha + \beta/2 + \gamma/2(1-\ell)(\phi_{00}
+\phi_{10}) + \gamma/2\ell(\phi_{01}
+ \phi_{11}) \quad \text{and} \quad
\overline{A}_{w^*}=n/2n = 1/2.
\end{eqnarray*}
And, 
\begin{eqnarray*}
\sum_{i=1}^{n}\frac{(A_i-\overline{A}_{w^*})^2}{P(A_i|L^*_i)}&=&\sum^{n^*_{00}}\frac{(-1/2)^2}{\sum_l[(1-\pi_l)P(L=l|L^*=0)]}\\
&+& \sum^{n^*_{01}}\frac{(-1/2)^2}{\sum_l[(1-\pi_l)P(L=l|L^*=1)]}\\
&+& \sum^{n^*_{10}}\frac{(1-1/2)^2}{\sum_l[\pi_lP(L=l|L^*=0)]}\\
&+& \sum^{n^*_{11}}\frac{(1-1/2)^2}{\sum_l[\pi_lP(L=l|L^*=1)]}\\
&=& 1/4 \times \sum_{i=1}^{n}1/P(A_i|L_i^*) = n/2.\\
\sum_{i=1}^{n}\frac{\E[(Y_i-\overline{Y}_{w^*})](A_i-\overline{A}_{\tilde{w}})}{P(A_i|L^*_i)}
&=& \sum^{n^*_{00}}\frac{\beta/4-\gamma/2\phi_{00}+\gamma/4(1-\ell)(\phi_{00}
+\phi_{10}) + \gamma/4\ell(\phi_{01}
+ \phi_{11})}{\sum_l[(1-\pi_l)P(L=l|L^*=0)]} \\
&+& \sum^{n^*_{01}}\frac{\beta/4-\gamma/2\phi_{01} + \gamma/4(1-\ell)(\phi_{00}
+\phi_{10}) + \gamma/4\ell(\phi_{01}
+ \phi_{11})}{\sum_l[(1-\pi_l)P(L=l|L^*=1)]}\\
&+& \sum^{n^*_{10}}\frac{\beta/4+\gamma/2\phi_{10} - \gamma/4(1-\ell)(\phi_{00}
+\phi_{10}) - \gamma/4\ell(\phi_{01}
+ \phi_{11})}{\sum_l[\pi_l P(L=l|L^*=0)]}\\ 
&+& \sum^{n^*_{11}}\frac{\beta/4+\gamma/2\phi_{11} - \gamma/4(1-\ell)(\phi_{00}
+\phi_{10}) - \gamma/4\ell(\phi_{01}
+ \phi_{11})}{\sum_l[\pi_l P(L=l|L^*=0)]}\\
&=& n(1-\ell)(\beta/4-\gamma/2\phi_{00}+\gamma/4(1-\ell)(\phi_{00}
+\phi_{10}) + \gamma/4\ell(\phi_{01}
+ \phi_{11}))\\
&+& n\ell(\beta/4-\gamma/2\phi_{01} + \gamma/4(1-\ell)(\phi_{00}
+\phi_{10}) + \gamma/4\ell(\phi_{01}
+ \phi_{11}))\\
&+& n(1-\ell)(\beta/4+\gamma/2\phi_{10} - \gamma/4(1-\ell)(\phi_{00}
+\phi_{10}) - \gamma/4\ell(\phi_{01}
+ \phi_{11}))\\
&+& n\ell(\beta/4+\gamma/2\phi_{11} - \gamma/4(1-\ell)(\phi_{00}
+\phi_{10}) - \gamma/4\ell(\phi_{01}
+ \phi_{11}))\\
&=& n/2(\beta(1-\ell) + \beta\ell
- \gamma(1-\ell)\phi_{00}
- \gamma\ell\phi_{01}
+ \gamma(1-\ell)\phi_{10}
+ \gamma\ell\phi_{11})\\
&=& n/2(\beta + \gamma(1-\ell)(\phi_{10} - \phi_{00}) + \gamma\ell(\phi_{11}-\phi_{01})
\end{eqnarray*}
In conclusion,
\begin{eqnarray*}
\E[\hat{\beta}] = \beta + \gamma(\phi_{10} - \phi_{00})(1-\ell) + \gamma(\phi_{11}-\phi_{01})\ell \quad \text{and} \quad
\E[\hat{\alpha}_{\mathrm{msm}}] &=& \alpha + \gamma/2 \times [2(1-\ell)\phi_{00} + 2\ell\phi_{01})]\\
&=& \alpha + \gamma \phi_{00}(1-\ell) + \gamma  \phi_{01}\ell.
\end{eqnarray*}

\begin{singlespacing}
\section{Illustration: sensitivity analysis of classification error in a confounding variable}
\end{singlespacing}\label{ap:sec:sensana}
To inform a sensitivity analysis, one needs to make an informed guess about the values for $p_0$, $p_1$. From the data, one can estimate $\ell$, $\omega$, $\pi^*_{0}$ and $\pi^*_{1}$. We calculate $\lambda, \pi_0$ and $\pi_1$ by using the data and the assumed $p_0$ and $p_1$. Since, 
\begin{eqnarray*}
\lambda = \frac{\ell - p_0}{p_1 - p_0}, \quad
\pi_0^* = \frac{\pi_0(1-p_0)(1-\lambda)+\pi_1(1-p_1)\lambda}{(1-\ell)}, \quad \text{and} \quad
\pi_1^* = \frac{\pi_0p_0(1-\lambda) + \pi_1p_1\lambda}{\ell},
\end{eqnarray*}
it follows that,
\begin{eqnarray}
\pi_0 = \frac{\pi_0^*(1-\ell)-\pi_1(1-p_1)\lambda}{(1-p_0)(1-\lambda)}, \quad
\pi_1 = \frac{\pi_1^*\ell-\pi_0p_0(1-\lambda)}{p_1\lambda}.\label{eq-app:pi_0pi_1}
\end{eqnarray}
By rewriting the expression for $\pi_1$ by using the expression for $\pi_0$, it follows that,
\begin{eqnarray*}
\pi_1 &=& \frac{\pi_1^*\ell-\pi_0p_0(1-\lambda)}{p_1\lambda}\\
&=& \frac{\pi_1^*\ell-\frac{\pi_0^*(1-\ell)-\pi_1(1-p_1)\lambda}{(1-p_0)(1-\lambda)}p_0(1-\lambda)}{p_1\lambda}\\
&=& \frac{\pi_1^*\ell-(\pi_0^*(1-\ell)-\pi_1(1-p_1)\lambda)\frac{p_0}{(1-p_0)}}{p_1\lambda}\\
&=& \frac{\pi_1^*\ell-\pi_0^*(1-\ell)\frac{p_0}{(1-p_0)}+\frac{(1-p_1)p_0}{(1-p_0)}\lambda\pi_1}{p_1\lambda}\\
&=& \frac{\pi_1^*\ell-\pi_0^*(1-\ell)\frac{p_0}{(1-p_0)}}{p_1\lambda}+\frac{(1-p_1)p_0}{(1-p_0)p_1}\pi_1\\
&=& \frac{\pi_1^*\ell-\pi_0^*(1-\ell)\frac{p_0}{(1-p_0)}}{p_1\lambda}+\frac{(1-p_1)p_0}{(1-p_0)p_1}\pi_1.
\end{eqnarray*}
Consequently,
\begin{eqnarray}
(1-\frac{(1-p_1)p_0}{(1-p_0)p_1})\pi_1 &=&  \frac{\pi_1^*\ell-\pi_0^*(1-\ell)\frac{p_0}{(1-p_0)}}{p_1\lambda},\nonumber\\
\pi_1 &=& \frac{\frac{\pi_1^*\ell-\pi_0^*(1-\ell)\frac{p_0}{(1-p_0)}}{p_1\lambda}}{\frac{(1-p_0)p_1 - (1-p_1)p_0}{(1-p_0)p_1}}\nonumber\\
&=& \frac{\pi_1^*\ell-\pi_0^*(1-\ell)\frac{p_0}{(1-p_0)}}{p_1\lambda}\times \frac{(1-p_0)p_1}{(1-p_0)p_1 - (1-p_1)p_0}\label{eq-app:pi_1}.
\end{eqnarray}
From model (\ref{eq-app:pi_1}) we now obtain an value for $\pi_1$, which we use to get obtain a value for $\pi_0$ from model (\ref{eq-app:pi_0pi_1}). The bias expressions (\ref{eq:biascm}) and (\ref{eq:biasmsm}) can subsequently be used to calculate bias in the ATE. 
\end{document}